\documentclass[twocolumn,prb,showpacs,10pt]{revtex4}
\usepackage{psfrag}
\usepackage{graphicx}
\usepackage{amsmath}
\usepackage{amssymb}
\usepackage{subfig}

\usepackage{color}
\usepackage{multirow}
\begin{document}
\title{Charge localisation on a redox-active single molecule junction and its influence on coherent electron transport}
\author{Georg Kastlunger and Robert Stadler} 
\affiliation{Department of Physical Chemistry, University of Vienna, Sensengasse 8/7, A-1090 Vienna, Austria \\ 
Email: robert.stadler@univie.ac.at}

\date{\today}

\begin{abstract}
For adjusting the charging state of a molecular metal complex in the context of a density functional theory description of coherent electron transport through single molecule junctions, we correct for self interaction effects by fixing the charge on a counterion, which in our calculations mimics the effect of the gate in an electrochemical STM setup, with two competing methods, namely the generalized $\Delta$ SCF technique and screening with solvation shells. One would expect a transmission peak to be pinned at the Fermi energy for a nominal charge of +1 on the molecule in the junction but we find a more complex situation in this multicomponent system defined by the complex, the leads, the counterion and the solvent. In particular equilibrium charge transfer between the molecule and the leads plays an importanty role, which we investigate in dependence on the total external charge in the context of electronegativity theory. 
\end{abstract}
\maketitle

\begin{section}{Introduction}\label{sec:intro}

Most studies in the vibrant field of single-molecule electronics focus on the low bias current flow through rather small benchmark molecules anchored to metal leads in ultrahigh vacuum (UHV) at very low temperatures. Under those restrictions the underlying electron transport problem is nowadays straightforwardly accessible to a computational treatment with a nonequilibrium Green's function (NEGF) approach~\cite{keldysh} in combination with a density functional theory (DFT) based description of the electronic structure of the separate and combined components of the junction, namely the leads and the scattering region~\cite{atk}$^-$~\cite{kristian}. This method allows for an atomistic interpretation of associated UHV experiments on such benchmark systems in a mechanical break-junction or scanning tunneling microscope (STM) setup~\cite{chavy,tour,lohneysen,ruitenbeek}, thereby contributing to a fundamental understanding of the dependence of the electronic conductance of the junction on the details of its structure within the boundary conditions of a low pressure and low temperature regime.

\begin{figure}
      \includegraphics[width=0.95\linewidth,angle=0]{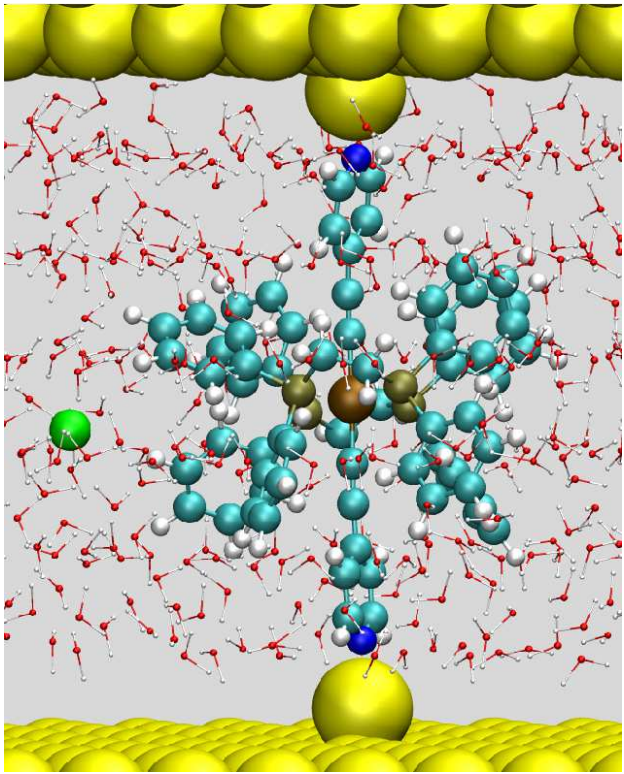}
        \caption[cap.Pt6]{\label{fig.complex}Geometry of the Ru(PPh$_2$)$_4$(C$_2$H$_4$)$_2$ bis(pyridylacetylyde) complex studied throughout our article bonded to ad-atoms on Au fcc (111) surfaces within an aqueous solvent and containing a Cl counterion.}
\end{figure}

For single-molecule junctions to be useful as molecular devices, however, their operability at room temperature is required and the presence of a solvent allows for electrochemical gating, which makes it possible to avoid the potentially destructive effect of the rather high local electric fields, which otherwise would be needed for inducing a larger current~\cite{Tim1}. Experimentally, these ambient conditions can be achieved with an electrochemical STM~\cite{Tim1,Tim2,Nichols1,Nichols2}, where the nano junction is an integral part of an electrochemical cell and the investigated molecules usually have a redox-active center with an oxidation state which can be regulated via gating~\cite{Tim1}. Depending on the setup as well as structural details of the system, two competing electron transport mechanisms have to be considered for a theoretical description of such experiments, namely electron hopping which is a thermally induced multiple step process and coherent tunneling which is the standard one-step phenomenon known from benchmark molecules without a redox-active center and relatively strongly coupled to metallic electrodes at temperatures close to 0 K. In both cases an atomistic description of the process under electrochemical conditions provides a formidable challenge for a DFT based theory. For the former, the difficulty lies in a simplified and compact but nevertheless sufficiently accurate description of the nuclear vibrations of the molecule and solvent which drive the electron flow. For the latter it becomes necessary to adjust the oxidation state of the redox active center in the scattering region and therefore deal with the issue of charge localization in a multi-component system, which is the topic we address in this article.

A correct description of localized charges is notoriously hard to achieve within a DFT framework, because the Coulomb and exchange parts of the interaction of an electron with itself do not cancel out exactly in a standard Kohn-Sham (KS) Hamiltonian and the corresponding self interaction errors (SIE) result in an artificial tendency towards delocalization~\cite{Perdew-Zunger,victor-paper,sanvito-victorpaper,graefenstein-victorpaper}. As has been shown recently, both for a continuum solvation model~\cite{Siegbahn} and for an explicit description of a periodic cell with its vacuum part filled up with H$_2$O molecules~\cite{Elvar}, a polar solvent has a screening effect on the Coulomb potential which reduces SIE and stabilizes localized charges within DFT. Another way to enforce localization is based on the generalized $\Delta$ SCF technique~\cite{Schiotz1,Schiotz2}, where an arbitrary integer value between 0 and 2 for the occupation number of a particular crystal eigenstate or linear combination of crystal orbitals can be defined as a boundary condition to the self-consistency cycles determining the electronic structure of a given system.

In our article we pursue both avenues for a study of the coherent electron transport through the Ru(PPh$_2$)$_4$(C$_2$H$_4$)$_2$ bis(pyridylacetylyde) complex in Fig.~\ref{fig.complex}, which we will often refer to  as just "the Ru-complex" in the following since it is the only system we investigate here and where for experiments in an aqueous solution with chlorine counterions the oxidation state of the redox active ruthenium atom can be switched between +II and +III by varying the electrochemical potential of the cell corresponding to an overall charge of 0 and +1 on the molecular complex, respectively. We chose this particular system because it was used in previous conductance measurements~\cite{Frisbie, Wang} as a monomer of chains - albeit with different anchor groups -  where it was found that depending on the chain length either coherent transport or electron hopping is observed~\cite{Frisbie}. In addition spectroscopic and quantum chemical studies on similar Ru complexes~\cite{Abruna, Humphrey, Touchard, Dixneuf, Heath} suggest that this molecular species offers the possibility to easily link two carbon-rich chains to each other for the formation of reversible redox systems~\cite{Dixneuf16_1,Dixneuf16_2,Dixneuf18} with distinct optical transition properties~\cite{Dixneuf_6, Dixneuf_19}, thereby serving as a starting point for the investigation of chains with multiple redox active centers~\cite{Dixneuf}. In contrast to Ref.~\cite{Frisbie} we use pyridil groups as anchors to the leads because they provide peaks in the transmission function, which are narrow enough to assume that a charge on the complex has an impact on the conductance but broad enough to avoid the Coulomb blockade regime~\cite{robert-21,robert-20,robert-6}.

Although reports of conductance calculations on redox-active complexes have been published before~\cite{Evers1,Evers2}, we believe our article to be the first DFT based study of coherent electron transport through such a molecular complex which explicitly investigates the influence of the formal oxidation state of its central metal atom on the resulting transmission function. There have been previous studies on the impact the solvent has on smaller benchmark molecules without a redox center~\cite{Sanvito-solvent,Lambert,Watanabe,Luo}, where some of them~\cite{Sanvito-solvent,Lambert} have found a "chemical gating" effect, i.e. a shift in the transmission function induced by the surrounding molecules, which was explained by dipole fields. We do not consider configurational fluctuations of the solvent molecules in our article, not only because of the high computational demands this would generate for our rather large junction but also because it would lead to fluctuations in the charge on the Ru complex where a main aim in this work is to keep it fixed and to study its influence in a systematic way.

It has to be stressed that by this restriction we neglect an important solvent effect, which would modify electron transport due to the related electron phonon coupling. While this effect is crucial for electron hopping -which is not the topic of this article- we believe our omission to be justified in the context of coherent tunnelling where the solvents main influence is of an electrostatic nature and the statistics for the positions of water nuclei should change the transmission fucntion and conductance of the junction only to a small extent.
The main electrostatic screening effect of the solvent in our calculations, namely the localization of the charge on the counter ion, can also be mimicked in a more technical way by fixing the charge on a Cl atom with the $\Delta$ SCF technique and in this article we compare the results of this approach with that of the explicit presence of the solvent.

The paper is organized as follows: In the next section we present transmission functions and conductances for the Ru complex at charging states of 0 and +1 (i.e. with the Ru atom in its formal oxidation state +II and +III, respectively), where in order to mimic the gate potential generating the +1 state in experiments, the counter charge is localized on a chlorine ion, and we assume that a Cl atom oxidizes the complex and thereby reduced to an anion. We do not suggest that this redox process necessarily takes place in the actual STM experiments but rather use it as a convenient tool to simulate the effect of electrochemical gating, namely charging the Ru complex in the junction, in our calculations. The two ways of reducing SIE as mentioned above, i.e. employing the generalized $\Delta$ SCF technique and introducing H$_2$O molecules explicitly as a solvent are used for making sure that the Cl atom is indeed charged with a whole electron in our setup. In Section~\ref{sec:projections} we investigate the shift in projected molecular eigenvalues with both methods in terms of the distribution of partial charges throughout the junction, which is a multi-component system in the sense that implementing the gate does not only involve the charge on the Ru-complex and counterion but also the gold leads and aqueous solvent can and do lose or gain fractions of electrons. For an analysis of this complex behaviour in Section~\ref{sec:charge} we start from cluster models within the simplified picture of electronegativity (EN) theory~\cite{en-theory}, and from their direct comparison with our full calculations on the junctions represented by Fig.~\ref{fig.complex} we derive the nature of the driving forces, which define the charge density distributions we observe. We conclude with a brief summary of our results.

\end{section}

\begin{section}{Electron transport calculations for the neutral and charged complex}\label{sec:transport}

All calculations of transmission probabilities T(E) in this article were performed within a NEGF-DFT framework~\cite{atk}$^-$~\cite{kristian} with the GPAW code~\cite{GPAW1,GPAW2}, where the core electrons are described with the projector augmented wave (PAW) method and the basis set for the KS wavefunctions can be optionally chosen to be either a real space grid or a linear combination of atomic orbitals (LCAO), and we opted for the latter on a double zeta level with polarisation functions (DZP) for all of our electron transport and electronic structure calculations. The sampling of the potential energy term in the Hamiltonian is always done on a real space grid when using GPAW, where we chose 0.18 \AA{} for its spacing and a Perdew-Burke-Ernzerhof (PBE)~\cite{PBE} parametrisation for the exchange-correlation (XC) functional throughout this article. 

Within NEGF the transmission function T(E) is defined by $T(E)= Tr(G_d\Gamma_L G_d^\dagger \Gamma_R)$ where $G_d = (E - H_d - \Sigma_L - \Sigma_R)^{-1}$ represents the Greens function of the device containing the self energy matrices $\Sigma_{L/R}$ due to the left/right lead, $\Gamma_{L/R} = i(\Sigma_{L/R}-\Sigma_{L/R}^\dagger)$ and H$_d$ the Hamiltonian matrix for the device region, which contains not only the Ru-complex but also 3-4 layers of the aligned Au surface on each side. Due to the rather large size of the central molecule (Fig.~\ref{fig.complex}), we had to use gold slabs with a 6x6 unit cell in the surface plane in order to ensure that neighbouring molecules do not interact. With the two Au ad-atoms directly coupling to the molecule (Fig. \ref{fig.complex}), the device region contains a total of 254 Au atoms in addition to the atoms of the complex itself and up to 64 H$_2$O molecules. As a consequence H$_d$ reached a size which was beyond our computational capabilities to be handled efficiently for electron transport calculations and therefore needed to be reduced. 

Since it is known that the solvent does not contribute to the peak structure in T(E)~\cite{Sanvito-solvent}, but instead adds a base-line conductance with a rather small energy dependence~\cite{Nitzan-review}, we cut out the lines and rows indexing H$_2$O basis functions in the matrix H$_d$, which we initially obtained from an electronic structure calculation for the full device region. In a second effort towards memory reduction we cut out very high- and very low-lying MOs from H$_d$ after sub-diagonalizing it with respect to molecular basis functions~\cite{rectifier12,rectifier23}, where we assumed that molecular eigenstates which are further than 5 eV apart from E$_F$ would have no effect on the conductance or on the transmission function on the much smaller energy range on which we show them. 

For ensuring overall charge neutrality in the unit cell of our device region which is a necessity for a charged junction when applying periodic boundary conditions for electronic structure calculations, the counter charge to the positively charged Ru-complex has to be an explicit part of the cell and we represent it by a Cl counterion. There are two methods we exploit in this article to overcome the SI problem, which leads to an artificial delocalization of otherwise localized charges in DFT: i) We make explicit use of the findings of other groups~\cite{Elvar,Siegbahn} that a polar solvent, H$_2$O in our case, stabilizes localized charges, because the solvation enthalpy and therefore also the total energy of the system become the more negative, the more point-like the charges on the solutes are distributed; ii) and we also employ the generalized $\Delta$SCF technique~\cite{GPAW2,Schiotz1,Schiotz2} which has been previously used as a feature of GPAW for a correct description of excitation processes in molecules adsorbed on surfaces~\cite{Elvar,Schiotz1} and of electron hopping between layers of oxides~\cite{Pawel1, Pawel2}.

In practical terms the first scheme starts with the relaxation of the nuclear positions of the isolated Ru-complex towards the convergence criterium of 0.02 eV/\AA{} for the average force. Then we add the Cl counterion with a fixed Ru-Cl distance of 7 \AA{} and embed the resulting system in a solvent shell of 46 molecules by making use of the graphical interface of the ghemical code~\cite{ghemical}, which places H$_2$O molecules in the cell with a high degree of artificial translational symmetry. In a next step we relax the nuclei of the system now comprising the complex, the counterion and the solvation shell in order to create a more natural distribution of water molecules, where hydrogen bonds create a network structure, but we keep the Ru-Cl distance constant as a boundary condition for avoiding hybridization between the Ru-complex and the chlorine ion, which is statistically unlikely in nature but might happen in our relatively small unit cell. During this relaxation process we regularly probe the charge distribution in the system. Once we achieve an one-electron charge on the Ru-complex, i.e. a formal oxidation state of +III on the Ru atom, we stop the relaxation and align the whole system between two gold fcc (111) surfaces with ad-atoms and the nitrogen of the pyridil anchors at a distance of 2.12 \AA{} for establishing the direct electronic contact~\cite{robert-20}. For this system we then calculate the transmission function as described above.

\begin{figure}
\includegraphics[width=0.95\linewidth,angle=0]{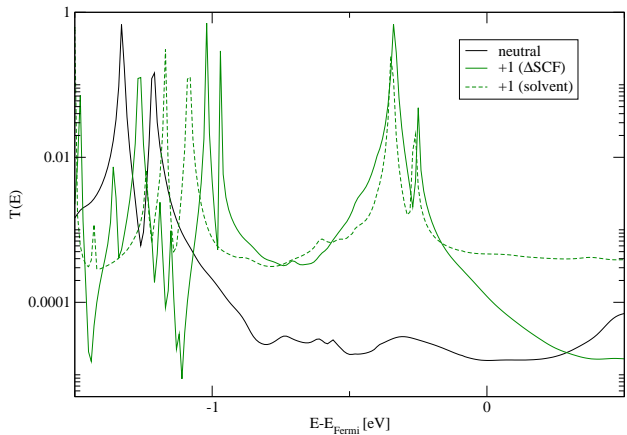}
\caption[cap.Pt6]{\label{fig.trans}Transmission function of the neutral Ru-complex (solid black line) and with a charge of +1 which was adjusted with two different methods, i.e. a) $\Delta$SCF (solid green line) and b) solvent screening (dashed green line). In both methods a Cl atom was used as a counterion to extract an electron from the Ru-complex. The kpoint sampling was performed on a 4x4x1 mesh for all three curves.}
\end{figure}

In our second approach based on the generalized $\Delta$ SCF method, we make use of its flexibility to define the spatial expansion of an orbital enforced to contain an electron as an arbitrary linear combination of Bloch states~\cite{Schiotz1,Schiotz2}. By extracting one electron from the system and inserting it into a predefined orbital in the beginning of every iteration step, the self consistency cycle progresses as usual but with the electron density of this particular orbital as a contribution to the external potential. In this way we can fix the electron occupation of the Cl counterion manually, which solves the self interaction problem implicitly and makes this method ideal for charge localization as needed in the present work. When applying this technique we chose the nuclear positions relaxed for the neutral complex aligned between the gold surfaces, where one counterion was added with one supplementary electron constrained to completely fill its p shell. This procedure also had the benign consequence that the calculation of T(E) was reduced significantly in terms of computational demand, because we do not need an explicit solvent here and therefore do not have to remove the respective states from the transport Hamiltonian.

\begin{table}
\begin{center}
 \begin{tabular}{|c|c|c|c|}
 \hline
&neutral molecule&		+1 ($\Delta$SCF)		& +1 (solvent)\\
\hline
G [G$_0$]&1.6$\cdot$10$^{-5}$&	1.2$\cdot$10$^{-4}$	&4.6$\cdot$10$^{-4}$\\
\hline
 \end{tabular}
 \caption{Conductance of the Ru-complex corresponding to the curves in Fig.~\ref{fig.complex} as calculated by NEGF-DFT and with the conductance quantum G$_0$ as its unit.}
\label{tab.conductances}
\end{center}
\end{table}

Fig.~\ref{fig.trans} shows the transmission function calculated for the neutral Ru-complex and with a positive charge put on the junction with the two methods described above. One would expect that the charged complex corresponding to a Ru-atom with an oxidation number of +III has a higher conductance than the neutral one (oxidation number +III) due to a supposedly half filled MO at the Fermi Level. While no Fermi level pinning can be observed in Fig.~\ref{fig.trans}, the conductance of the +1 state is indeed distinctly higher than that of the neutral junction as shown in Table~\ref{tab.conductances} but the respective numbers obtained from the two methods for applying the charge differ by a factor of four.  

The main reason for this disagreement is illustrated by Fig.~\ref{fig.trans}, where we find that the incompleteness in decoupling the H$_2$O orbitals from the transport Hamiltonian -conceding that LCAO basis functions located on specific atoms also contribute to the description of their surrounding- creates a "transmission baseline" which fits the behaviour previously investigated in theoretical studies of the conductance of water~\cite{Nitzan-review}, and is absent in the $\Delta$SCF calculations. In this line of argument, the difference of the transmission function and conductance for the +1 state calculated with $\Delta$ SCF and solvent screening is caused by the solvent retaining some presence in one of the transport Hamiltonians because electrons in the solvent are to some extent described by basis functions localized on the complex and therefore contribute to the transport.

\end{section}

\begin{section}{Charge density distribution and its impact on the projected MO eigenenergies}\label{sec:projections}

\begin{figure}
   \includegraphics[width=1.0\linewidth,angle=0,clip=True,trim=0cm 0cm 0cm 0cm]{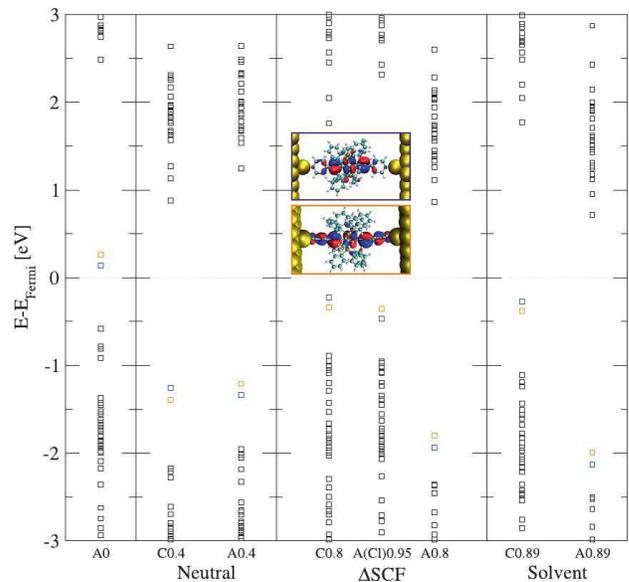}
  \caption[cap.Pt6]{MO eigenvalue spectrum of the device region, where the spatial shape of the HOMO and HOMO-1 are shown as insets, where the C panels are obtained from a subdiagonalization of the transport Hamiltonian and the A panels from a vacuum level alignment of the isolated molecule and leads. The numbers in the panel descriptions refer to the charge on the complex, where further technical details are described in the main text.
  }
  \label{fig.mo}
\end{figure} 

\begin{table*}
\begin{center}
 \begin{tabular}{|c|c|c|c|c|c|c|}
\hline
				&	Au	&	H$_2$O	&	Cl	&	H$_2$O+Cl	&	Ru complex	&	Ru\\ 
\hline
	Neutral complex	&	0.39	&	--	&	--	&	--		&	-0.43		&	-0.21	\\
\hline
	$\Delta$SCF	&	-0.16	&	--	&	0.94	&	0.94 (0.97)	&	-0.80 (-0.97)	&	-0.25 (-0.35) \\
\hline
Solvent	(1Cl/46 H$_2$O)	&	-0.21	&	0.37 (0.28)&	0.71 (0.70)&	1.08 (0.98)	&	-0.90 (-0.98)	&	-0.24 (-0.33) \\
\hline      
\end{tabular}
\caption[cap.Pt6]{Distribution of the partial charges in the junction as calculated from a Bader analysis for the neutral complex and the complex with one positive charge applied by fixing the counter charge on a Cl ion with $\Delta$SCF and solvent screening, respectively, where numbers from calculations without a Au slab are also shown in parantheses for comparison. All values are given in fractions of electrons.}
\label{tab.charges_all}
\end{center}
\end{table*}

\begin{table}
\begin{center}
 \begin{tabular}{|c|c|c|c|c|}
\hline
                &    \multicolumn{2}{|c|}{Uncorrected}    &    Solvent  &    $\Delta$SCF\\
\cline{2-5}
                    &    B3LYP        &    PBE &    PBE            &    PBE\\
\hline
Counter charge &    0.41        &    0.42     &    0.98            &    0.97 \\
\hline
\end{tabular}
\caption[cap.Pt6]{Partial charges on a Cl ion (and if applicable also on the solvent) sharing the same cell with the complex in absence of the Au leads in units of fractions of an electron. Both the solvent screening method and $\Delta$ SCF generate the correct result of one electron, while B3LYP underestimates charge localisation in the same way as PBE.}
\label{tab.charges_noAu}
\end{center}

\end{table}

 \begin{figure}
  \includegraphics[width=0.48\textwidth]{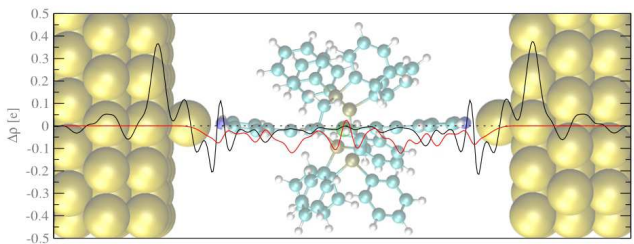}
  \caption[cap.Pt6]{Charge density difference between the coupled system and the isolated complex and gold slab (black curve), and between the isolated complex in its charged and neutral state (red curve), where pseudo densities in terms of the PAW formalism have been used for the densities, in order to eliminate artificial peaks near the nuclei.}
  \label{fig.elden}
 \end{figure}

In order to understand the peak structure in Fig. \ref{fig.trans} in more detail we now study the electronic structure of the junction by investigating the electronic states of the device in terms of the molecular eigenenergies and their shape. Since the coupling of the Ru-complex to the Au surface leads to a hybridisation of the respective electronic states, it is necessary for the projection of molecular eigenvalues localized on the Ru-complex from the Hamiltonian matrix to eliminate their coupling to the surface states in a subdiagonalization procedure~\cite{rectifier12,rectifier23}. The MO-eigenvalue distributions obtained in this way are shown in Fig. \ref{fig.mo}. 
The MO eigenenergies are calculated by decoupling the basis functions localized on the molecule from that of the surface states with a subdiagonalization of the transport Hamiltonian for the neutral complex for panel C0.4, and for a complex with a charge of +1 applied by $\Delta$SCF and the solvent screening method for panels C0.8 and C0.89, respectively. The energies in panels A0, A0.4, A0.8 and A0.89 result from vacuum level alignment of separate calculations for the Ru-complex and the Au slab, where the numbers in the panel labelling refer to a positive charge of that size on the complex. For panel A(Cl)0.95 a chlorine atom is added to the Ru-complex for the alignment.
By inspecting the shape of the two relevant orbitals for coherent transport through the Ru-complex in both charging states, namely the HOMO and HOMO-1 which we show as insets in Fig. \ref{fig.mo}, we find that both MO's are characterized by a conjugated $\pi$-system, which is delocalized over the whole bridge of the complex and their respective energies match with the double-peak structure in the transmission function in Fig. \ref{fig.trans}. While the HOMO-1 in Fig. \ref{fig.mo} has a high localization at the interface region, the HOMO does not, which explains the relative proportions of the widths of the two merged peaks in Fig. \ref{fig.trans}.

For very weak coupling between the leads and a molecule one would expect that charging the molecule to its +1 state extracts one electron from the complex's HOMO leading to a SOMO which by definition is situated at the Fermi energy E$_F$. In the composite junction we investigate in this article, however, where the degree of electronic coupling is intermediate and we can only obtain molecular orbitals by projecting them out of lead/complex hybrid states via a dehybridization procedure, the situation is less clear cut and in Fig. \ref{fig.mo} we find the HOMO always below the junctions Fermi level, which is mostly defined by the leads due to their metallic character and the large number of gold atoms in the device region. The key for understanding the peak positions in the transmission function and the Fermi level alignment of the corresponding MOs in such a scenario lies in understanding the zero bias charge transfer as has been demonstrated in Refs.~\cite{robert-16,robert-13,robert-5} for bipyridine and other similarly small organic molecules. The present case, however, is more difficult because here we have to deal with a four component system containing the Ru-complex, the Cl ion, the solvent and the leads, where for a detailed charge density distribution analysis we use the Bader method for the definition of the electronic charges belonging to particular nuclei~\cite{bader,bader2} in the following. 

In Table \ref{tab.charges_all} we present the charge distribution for both the neutral and charged junction, where values from separate simulations for the Ru-complex without Au leads but for the charged case including the counterion and solvent are given in parentheses for comparison but are also highlighted in Table \ref{tab.charges_noAu} and there compared with values calculated with the hybrid functional B3LYP. In the absence of the Au surface the charge values on the Ru-complex can be adjusted rather precisely with both applied charge localization methods with the only difference between them that with solvent screening 28\% of the negative counter charge is found on the solvent and $\Delta$SCF by definition puts a whole electron on the chlorine. We also illustrate in Table \ref{tab.charges_noAu} that a small admixture of Hartree Fock exchange as it is contained in the B3LYP functional with the aim of reducing SI effects does not necessarily help to obtain the physically correct charge localisation as has been discussed by one of us in the context of electron coupling in a recent article~\cite{victor-paper} and the functional is impractical for a treatment of the whole junction in terms of computational expediency.

While gold creates a new reference energy for the molecular eigenstates, it also plays the role of an electron donor or acceptor, meaning, that it can accept charge from both the complex and the counterion/solvent system. We also note in this context, that for pyridil anchors on gold surfaces Pauli repulsion leads to an electron depletion on the complex which lowers its eigenstates energetically~\cite{robert-16}. This is exactly what we also find for the neutral complex in the composite junction here, where it loses electrons to the Au surface and Fig. \ref{fig.elden} shows that the charge transfer happens mostly at the interface, while the rest of the junction is not contributing to it in a significant way, while for the charged junction the gold bulk absorbs some of the positive charge as displayed in Table \ref{tab.charges_all}. 

It is a delicate question wether this latter charge absorption is due to SI artefacts in the calulations or a realistic result for the investigated system. While we deal explicitly with the SI error for the charge localisation on the chlorine counterion, the charge distribution between the Ru-complex and the gold slab is not necessarily strongly localised anywhere. The Ru atom is embedded into the complex by rather strong covalent bonds with its carbon ligands and as a consequence it contains only a fraction of a positive charge in both the neutral and charged complex (i.e. for its formal oxidation numbers +II and +III) and regardless of whether the complex is attached to the surface or not as can be seen from the numbers in Table \ref{tab.charges_all}. Also the electronic coupling at the interface is of intermediate strength, as indicated by the rather broad peak shape in the transmission functions. This does not contradict with the fact, that the bonding between the pyridil anchor group and gold atom is rather weak~\cite{robert-20}, because in the case of Pauli repulsion the coupling with filled MO's produces bonding and antibonding states~\cite{robert-16}. So the charge distribution we find in Table \ref{tab.charges_all} could be physically correct, although it is not what one would attribute to the system when writing down its redox equations. For investigating the issue whether the charge distribution in the junction is realistic further, we employ electronegativity theory in the next section, where we reduce the complexity of the investigated four component system by replacing the chlorine ion and solvent by an external charge for our analysis.

At this point we just use the partial charges computed with Baders method and given in Table \ref{tab.charges_all} for analyzing the contributions defining the projected MO eigenenergies in Fig.~\ref{fig.mo} in the way established in Ref.~\cite{robert-16}. In the panel A0 we align vacuum potentials between the isolated Au slab and the isolated Ru complex without any charging of the components, which results in the HOMO and HOMO-1 being energetically higher than the Fermi level of the gold leads. If we consider the changes in the respective vacuum potentials due to the negative charge on the Au slab (+0.39 electrons) and the positive one on the complex (-0.43 electrons), we arrive at the level positions given in panel A0.4 with the HOMO and HOMO-1 well below E$_F$, which almost exactly match with the projections from the composite junctions which are also shown as C0.4. This good agreement is somewhat surprising given that while the Paul repulsion effect depletes electrons mainly from the pyridil anchor groups of the Ru-complex, a partial charge externally put on the isolated complex is distributed evenly because it is achieved by emptying the HOMO as can be seen by comparing the black and red curves in Fig. \ref{fig.elden}. The situation becomes more complicated for the charging state +1 of the junction, where there is an apparent mismatch between MO projections from the composite system, panels C0.8 and C0.89 for $\Delta$SCF and solvent screening, respectively, and their analogons from the vacuum alignment of the separated Au slab and Ru-complex, panels A0.8 and A0.89, where the partial charges from Table \ref{tab.charges_all} have been applied externally. 

Although it is natural that A0.8 and A0.89 exhibit lower eigenergies of MOs than A0.4 due to the increased binding of electrons in more strongly positively charged molecules, the HOMO has to be close to E$_F$, i.e. within the range of the Fermi width, because it is partially emptied for charging state +1, which is indeed the case for the projections in panels C0.8 and C0.89. The solution to this conundrum can be found when considering the role of the counterion which also has an influence on the vacuum potential if now the Ru-complex and the chlorine are considered to be one component in the alignment process with the Au slab being the other one. This scenario is depicted in panel A(Cl)0.95, where we perform the level alignment starting from a calculation with a chlorine charged with an electron by $\Delta$SCF and extracting the counter charge from the complex as the molecular component. Unfortunately we can define our constraints within $\Delta$SCF only for integer charges but a hypothetic A(Cl)0.8 would result in slightly higher MO eigenenergies compared to A(Cl)0.95 and therefore be in perfect agreement with C0.8 in Fig.~\ref{fig.mo}. The distinct rise in energies going from A0.8 to A(Cl)0.95 is intuitively clear, because we are replacing the vacuum potential of a strongly positively charged component with that of a strongly polarized but overall neutral one. We note that in all cases HOMO and HOMO-1 switch their respective energetic positions as indicated by the colors used in Fig.~\ref{fig.mo}, which can be readily explained by their different localization patterns at the interface which we referred to at the beginning of this section.

\end{section}

\begin{section}{Interpretation of the charge distribution in terms of electronegativity theory}\label{sec:charge}

In order to find explanations for the charge density distributions described in the last section, we now analyze the junction in terms of electronegativity theory following the concepts of Parr and Pearson~\cite{en-theory}. The key quantities in this approach are the electronegativity $\mu$ and the hardness $\nu$, where the first is based on Mulliken's definition of electronegativity\cite{mulliken}, i.e.

\begin{equation}
 \mu=\left(\frac{\partial E}{\partial N}\right)_q = \frac{I + A}{2}
 \label{eq:mu}
\end{equation}

and the latter is defined as

\begin{equation}\label{eq:nu}
 \nu = \frac{1}{2}\left(\frac{\partial^2 E}{\partial N^2}\right)_q = \frac{I-A}{2}
\end{equation}

with I being the ionisation potential, calculated as the total energy difference of the N and N-1 system, and A the electron affinity, defined as $ E(N+1) - E(N)$. 

When two different systems are brought into contact the charge transfer from one to the other can be calculated as

\begin{equation}
 \Delta N = \frac{\mu_2 - \mu_1}{2(\nu_1+\nu_2)}
 \label{equ:dN}
\end{equation}

where both the electronegativities and hardnesses of the separate components have an impact on the amount of charge transfer between them~\cite{en-theory}.

\begin{table}[t]
 \begin{tabular}{|c|c|c|c|c|}
\hline
    \multirow{2}{2cm}{\centering\# of gold atoms}	&  \multicolumn{2}{|c|}{Starting charges}	& \multirow{2}{1cm}{\centering$\Delta$N}	& \multirow{2}{1cm}{\centering$\overline{\Delta N}$}\\
\cline{2-3}
					&	Ru-complex	&  Au-cluster		&		&		\\
\hline
\multirow{5}{1cm}{\centering2\\\includegraphics[width=1.3cm,clip=true,trim=3cm 4cm 1.2cm 4cm]{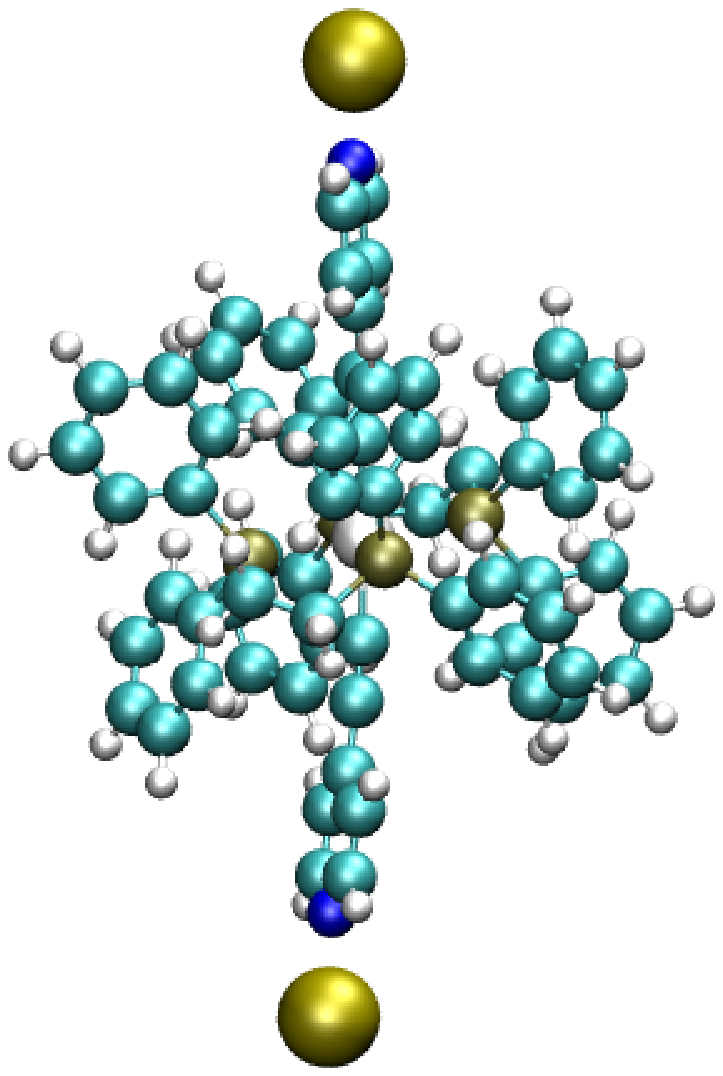}}	&	+1		&	0		&	-0.69	&	\multirow{2}{1cm}{\centering-0.66}	\\
					&	0		&	+1		&	-0.64	&				\\
\cline{2-5}
					&	+2		&	0		&	-1.28	&	\multirow{3}{1cm}{\centering-1.17}	\\
					&	+1		&	+1		&	-1.29	&				\\
					&	0		&	+2		&	-0.95	&				\\
\hline
\multirow{5}{1cm}{\centering254\\\includegraphics[width=1cm,clip=true,trim=0cm 0cm 0cm 0cm]{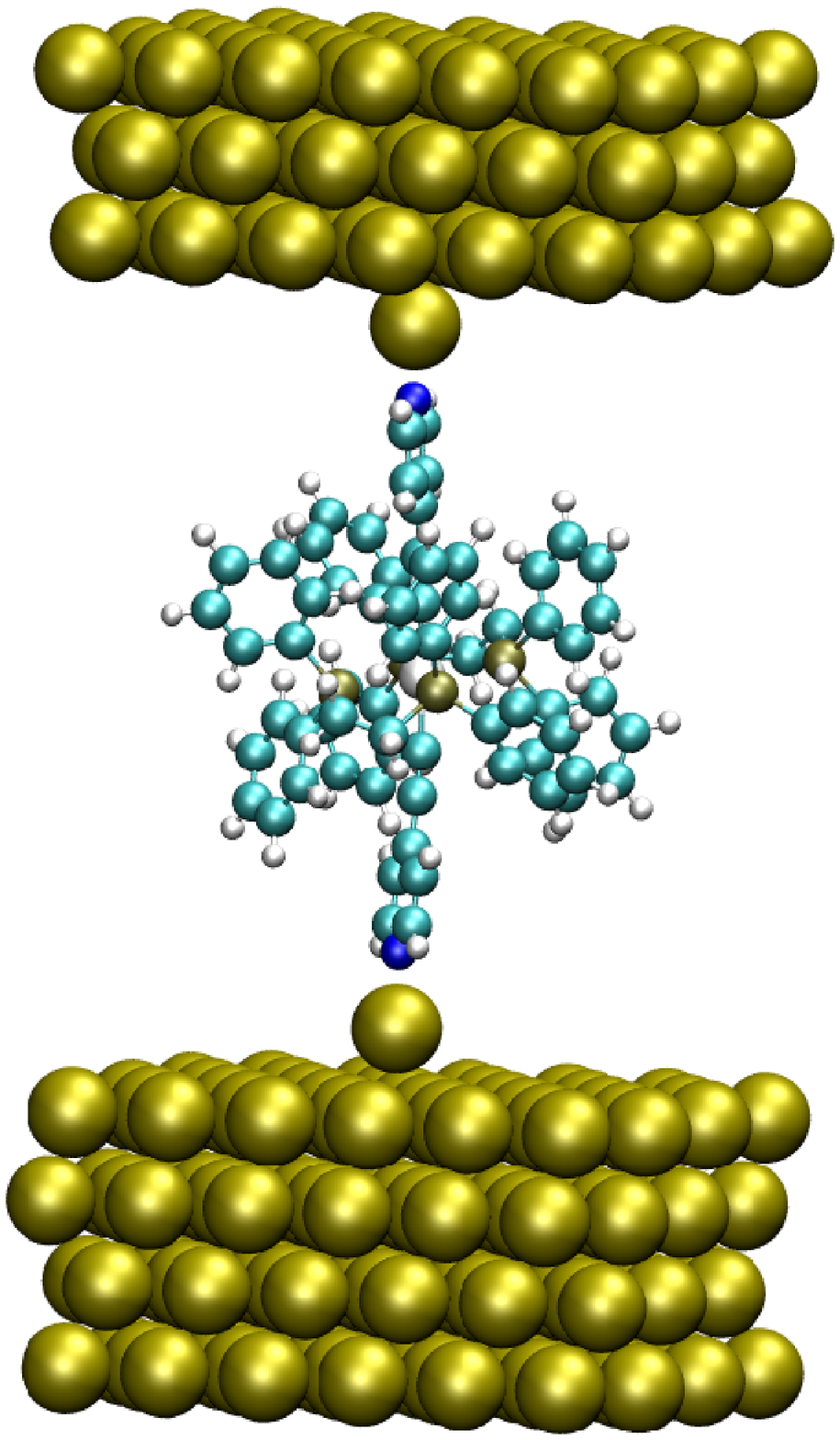}}	&	+1		&	0		&	-0.24	&	\multirow{2}{1cm}{\centering-0.29}	\\
					&	0		&	+1		&	-0.34	&				\\
\cline{2-5}
					&	+2		&	0		&	-0.52	&	\multirow{3}{1cm}{\centering-0.51}	\\
					&	+1		&	+1		&	-0.50	&				\\
					&	0		&	+2		&	-0.50	&				\\
\hline
 \end{tabular}
\caption[cap.Pt6]{Illustration of the statistics in our EN theory predictions for charged states, which arises from the possibility of different initial charge configurations on the subsystems before they are brought into contact. The point of reference for $\Delta$N in this table is the Ru-complex in its charging state 0.}
\label{tab.statistics}
\end{table}

 \begin{figure}   
 \includegraphics[width=1.1\linewidth,angle=0,clip=true,trim= 1cm 2cm 0cm 4cm]{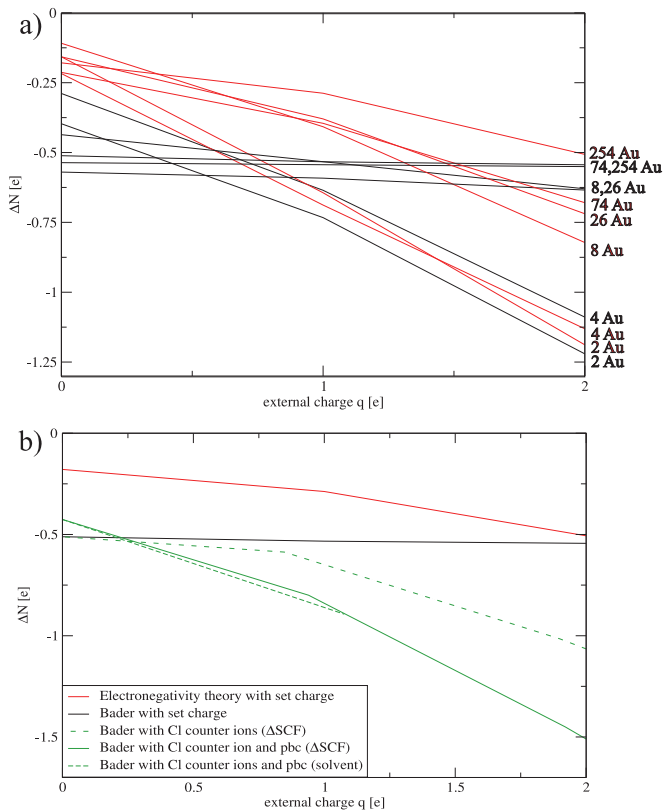}   
 \caption[cap.Pt6]{\label{fig.elneg}Electron loss on the complex when brought into contact with Au clusters of varying size and an external charge of up to +2$|e|$ is applied. Panel (a) shows the values predicted from electronegativity theory (red) and from calculations where the complex is coupled to gold clusters in a composite system and the charge distribution analyzed with the Bader analysis (black). In panel (b) the $\Delta$N values from these two sets of model calculations are compared with calculations of the device region, where the external charge was imposed as counter charge localized on Cl ions with and without periodic boundary conditions (pbc) which are shown as solid and dashed green lines.}
 \end{figure}     

The ionisation potential I and the electron affinity A are commonly defined for the neutral state of the individual subsystems, but as shown by Balbas et al.~\cite{Alonso}, their role of defining the electronegativity and hardness is also valid for ions, which allows us to describe also the charge distribution in the junction with a charging state of +1 in terms of EN theory. As discussed in the previous sections we fixed the charges on the counterion and solvent manually, and therefore in this section we are mostly interested in understanding the charge distribution between the Au slab and the Ru-complex. For this purpose we adjust their respective charging states by putting an external charge q on the subsystems in separate calculations without periodic boundary conditions where the charge in the simulation cell can be defined by the total number of electrons without having to worry about electrostatic interactions with neighbouring cells. The definition of $\mu$ and $\nu$ in Eqns. \ref{eq:mu} and \ref{eq:nu} as functions of such an external charge q is unusual, but not in contradiction to the basic assumptions of EN theory. 

It requires, however, some statistics for taking into account the possible starting points for the charge transfer. In the case without external charge only one such initial electron configuration of the components has to be dealt with, i.e. a neutral gold slab and a neutral complex. Raising the external charge to +1$|e|$ allows for two different starting points for the charge transfer, namely Ru-complex$^{+1}$/Au$^0$ and Ru-complex$^0$/Au$^{+1}$. In principle the calculation of $\Delta$N for both should lead to identical predictions for the final charge distribution in the composite system with a total charge of +1, but imperfections of our DFT based total energy calculations such as SI errors and the approximative nature of the XC functional lead to deviations, as shown in Table \ref{tab.statistics}. Averaging $\Delta$N over all possible integer configurations should provide an improvement with regard to such errors. Hereby special emphasis has to be put on the reference point for $\Delta$N, i.e. the subsystem with the index 1 in equation \ref{equ:dN}. Since the Ru-complex and the gold slab enter this equation at charged states, the calculated $\mu_i$,$\nu_i$ and $\Delta$N are also referring to these charged states. In order to obtain the change of electrons relative to the neutral subsystems the related integer charges therefore have to be subtracted. 

For understanding the role of the size of the gold slab for the charge distribution we model the gold component in our EN theory analysis with clusters of different sizes, starting from the adatom and reaching up to the full gold surface used in the junction, as shown in Fig. \ref{fig.elneg}, where we computed the electronegativities and hardnesses for charging states from 0 to +2 for each cluster size in a setup without periodic boundary conditions and calculated $\Delta$N averaged over initial electron configurations as described above. Although only charging states 0 and +1 correspond to the experimentally relevant oxidation states for the Ru-atom +II and +III, respectively, we nevertheless go to higher positive charges in this study in order to investigate the distribution between lead surface and metal complex in more general and systematic terms. In Table \ref{tab.statistics} we show the related statistical spread for the smallest and largest of our cluster sizes. Although we find that the deviations increase both with the external charge and the size of the Au cluster, their overall values are reasonably small indicating that our predictions for $\Delta$N from EN theory are not particularly limited in their accuracy by SIE or our choice of XC- functional.

For building a bridge between the predictions for the charge distribution from EN theory and the actual ones we find in the periodic systems we use as device regions in the transport calculations, we also performed cluster calculations, containing both subsystems. The charge distribution in the resulting cluster cells were analyzed according to Bader~\cite{bader,bader2}, where we imposed external charges for varying the charging state as we did for the subsystems for the EN predictions. The appeal of this intermediate step towards the periodic system calculation is that it allows us to distinguish between effects which come from electronegativity differences of the components, others which find their origin in the spatial polarisation of the subsystem, when they are actually brought into contact in a given geometry~\cite{robert-13,Nalewajski} and finally those related to the particular method we employ for adjusting the charging state.

As shown in Fig. \ref{fig.elneg}a the results from the EN prediction and the Bader analysis of the composite systems in dependence on the Au cluster size differ. In this comparison when we apply EN theory,  the charge transfer is slightly underestimated for an external charge q=0 $|e|$. Raising q to finite values leads to an overestimation of $\Delta$N with respect to the Bader analysis for the composite system. The deviation at high external charges is small for less then four gold atoms on both junction sides, but increases with the Au cluster size. 

Fig. \ref{fig.elneg}b puts a different perspective on these qualitative differences when we compare the charge distributions obtained from both EN theory and the Bader analysis of the cluster calculations with the results for the periodic device region (see also Table \ref{tab.charges_all}). While in the latter case the charge on the molecule increases almost linearly with the counter charge, the two models do not predict this behaviour for a cluster of 254 gold atoms. The reason can be found in the details of the charging state definition, where for the model calculations an external charge is imposed, which is distributed homogeneously, and for the subsystem the introduced charge delocalizes over all the atoms in the cluster leading to just a minor rise of its electronegativity with increasing q. This is a consequence of the hardness, as the derivative of the electronegativity (see Eqns. \ref{eq:mu} and \ref{eq:nu}) becoming smaller with cluster size. On the other hand the energy needed to extract an electron from the much smaller Ru-complex increases strongly with its charging state compared to the gold. As a consequence the external charge is mostly absorbed by the Au cluster, leading to rather modest charging of the molecule with an increasing external charge in the cluster models. 

\begin{figure}[t]
  \includegraphics[width=8cm,clip]{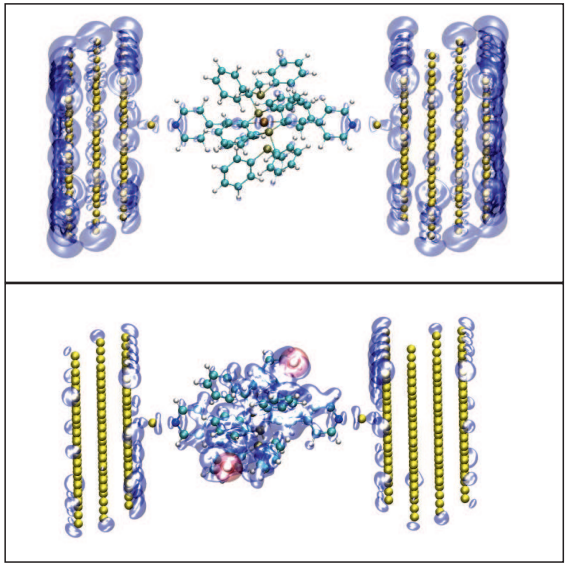}
  \caption[cap.Pt6]{Electron density difference between the Ru-complex in charging states +2 and 0, where the charge was put on the cluster by an external charge (upper panel) or Cl ions , with the negative counter charge localized by $\Delta$SCF (lower panel). In both cases we show results from cluster calculations with an isovalue threshold of 2*10$^{-4}$ e, where a loss of ele///ctronic charge is depicted in blue and a gain in red colour.}
  \label{fig.charge_loc}
\end{figure}

If on the other hand we adjust the charging state also in the composite cluster calculations in the same way we did for the periodic cells, namely by localizing the counter charge on a chlorine ion the situation changes, as can be seen from the dashed green curve in Fig. \ref{fig.elneg}b. Instead of a globally defined external charge we now have one or two point charges of opposite sign situated around the cluster. As a consequence a local Coulomb attraction term makes a localization of the positive charge on the Ru-complex and the Au surface rather than the bulk regions more favourable. Fig. \ref{fig.charge_loc} shows the charge density difference  between the +2 state and the neutral junction, for the charging state defined by an external charge (upper panel) and by chlorine atoms with charge localization enforced by $\Delta$SCF (lower panel). Without counterions the introduced positive charge is localized mostly on the gold atoms in the leads. Due to the non periodic setup of the cell fractional positive charges propagate to the outward pointing surfaces of the gold because of their mutual repulsion. If the charging state is defined by chlorine counterions on the other hand, the introduced positive charge is mostly localized on the Ru-complex and the lead surface because it is attracted by the counterions. Fractions of positive charge are, however, still localized on the outer parts of the gold bulk, since they are not hindered by the presence of a neighbouring cell in a non periodic setup and Fig. \ref{fig.elneg}b shows that therefore periodic boundary conditions even increase the positive charge on the Ru-complex region. 

\end{section}

\begin{section}{Summary}\label{sec:summary}

The aim of this article was the description of coherent electron transport through a single molecule junction containing a redox active center with an emphasis on its charging, a scenario which to our best knowledge has never been studied on an ab initio level before. A correct description of the charge distribution within DFT is essential in this context and we applied two independent methods for correcting the self interaction error, namely solvent screening and $\Delta$SCF, where in both cases the counter charge is localized on a Cl ion, where this setup is meant to mimic the effect of a gate in an electrochemical STM setup. We found that the actual charge on the Ru complex in a charging state of +1 (i.e. corresponding to a formal oxidation state of +III of the Ru-atom) is smaller than one, when it is coupled to a gold surface, which might indeed be realistic since some of the charge can be absorbed by the leads. In order to investigate this issue we made predictions for model systems of varying size of the gold component within electronegativity theory, which we supplemented with cluster calculations. This analysis led us to the conclusion, that some part of the charge should indeed be absorbed by the leads, but most of it remains on the complex due to Coulomb attraction, where the vicinity of the localized charge on the counterion has a stabilizing effect. Therefore, we assume that the charge distributions we find in our calculations for the device region are realistic in physical terms.

\end{section}


\begin{acknowledgments}
G.K. and R.S. are currently supported by the Austrian Science Fund FWF, project Nr. P22548. We are deeply indebted to the Vienna Scientific Cluster VSC, on whose computing facilities all calculations presented in this article have been performed (project Nr. 70174) and where we were provided with extensive installation and mathematical library support by Markus St\"{o}hr and Jan Zabloudil in particular. We gratefully acknowledge helpful discussions with Elvar \"{O}. J\'{o}nsson, Pawel Zawadzki, Marcin Dulak, Karsten W. Jacobsen, Kristian S. Thygesen, Victor Geskin and Tim Albrecht.
\end{acknowledgments}


\bibliographystyle{apsrev}

\end{document}